\documentclass[12pt]{article}
\usepackage{amsmath}
\usepackage{amssymb}
\usepackage{graphicx}
\usepackage{bm}
\usepackage{upgreek}
\begin{document}
\title{\bf \Large Teleparallelism by inhomogeneous dark fluid}

\author{Ertan G\"udekli$^1$\footnote{Email: gudekli@istanbul.edu.tr}, Aizhan  Myrzakul$^{2}$\footnote{Email: a.r.myrzakul@gmail.com} and Ratbay  Myrzakulov$^{3}$\footnote{Email: rmyrzakulov@gmail.com} 
\\ \textit{$^1$Department of Physics, Istanbul University, Istanbul, Turkey}
\\ \textit{$^2$Theoretical Physics Group, Department of Physics, }\\
\textit{Imperial College London, London SW7 2AZ, United Kingdom}
 \\ \textit{$^3$Eurasian International Center for Theoretical Physics,}  \\ \textit{Eurasian National University, Astana 010008, Kazakhstan} }

\date{}
 \maketitle


 \renewcommand{\baselinestretch}{1.1}

 \begin{abstract}
 
In this paper, we investigate $f(T)$ cosmology and find an exact solution for $f$ which gives a Little Rip cosmology. Also, considering accelerating cosmology with dark matter, the time-dependent solution is found. For these cases, by using solutions obtained from $f(T)$ gravity we find expressions for $\omega$ and $\Lambda$ defined as time functions via equivalent description in terms of inhomogeneous fluid. This puts the question: which theoretical model describes the observational cosmology? 
  
 \end{abstract}

\bigskip
\section{Introduction}

Recent studies indicate that our universe is expanding with acceleration, and it entered this phase recently. There are various models describing the acceleration and one of the most common is modification of Einstein-Hilbert action with a scalar curvature, where the expansion is mainly due to alternative action [1]. However, the application of this theory turns out to be difficult what is primarily associated with a complicated form of the corresponding field equations. Therefore, it is necessary to study the other descriptions for the expansion of the universe. One of the interesting type of the modified theory is $f(T)$ gravity, which was proposed recently. Beyond general relativity, "teleparallelism" with the Weitzenbock connection is considered as a gravitational theory \cite{f(T)}. This theory is described by the torsion $T$ and not by the curvature $R$ defined by
the Levi-Civita connection. Moreover, we show that such
$f(T)$ model also admits the dark energy epoch of the Universe without resorting to DE \cite{R1}.

According to the recent study, it was found that the equation of state parameter of dark fluid is about -1(equal, more or less), which corresponds to the dark energy density which increases monotonically with time. This shows that all the new gravity models are closely consistent with gravity theory in which the effective equation of state parameter is about -1. There are several interesting possible scenarios such as Big Rip [3], Little Rip [4], and
Pseudo-Rip [5] models which describe the fate of the
universe. In these models, the dark energy density is taken as a monotonically increasing function [6].
In this work we study Little Rip cosmology and other accelerating cosmology in $f(T)$ gravity. We show that it maybe equivalently described as the one due to inhomogeneous fluid [7].

The plan of the paper is as follows.
In Sec. 2, we give  the basic facts from the theory of the $f(T)$ gravity.  
In Sec. 3, we investigate the  accelerated cosmology with dark matter. 
Next, we study the  mimicking of $f(T)$ model by inhomogeneous fluids  in Sec.4.  In Sec. 5 
we consider some examples of integrable models of $f(T)$ gravity. 
Finally, we give our conclusions in Sec. 6.

\bigskip

\section{$f(T)$ cosmology}

The modified teleparallel action describing $f(T)$ gravity reads as
  
\begin{equation}
 I=\int d^4 x\sqrt{-g}[f(T)+L_m],  
 \end{equation}
where $L_m$ is the Lagrangian matter. An explicit expression for the scalar torsion $T$ is defined from the Einstein equation
  
\begin{equation}
 R_{\mu\nu}-\frac{1}{2}g_{\mu\nu}R=8\pi{G}T_{\mu\nu}\,\label{2.2}\\
 \end{equation} 
by substituting into this flat Friedman-Lemetre-Robertson-Walker metric
 
\begin{equation}
 ds^2=dt^2-a^2(t)dl^2,  
 \end{equation} 
where $a$ is the scale factor. Then the Friedmann equation takes the form
 
\begin{equation}
 H^2=\frac{k^2}{3}\left(\rho_{DE}+\rho_{DM}\right),
 \end{equation}
where $\textit{k}=8\pi{G}$, $\rho_{DE}$ is energy density of dark components, $\rho_{DM}$ is energy density of dark matter, $\textit{H}=\frac{\dot{a}}{a}$ is a Hubble parameter. It follows from (4) that $T=-6H^2$. Since the energy density of dark components  is much bigger than the density of dark matter, we can neglect the second term in Eq. (4).   
The density and the pressure of the dark  components can be written as [8]:

\begin{equation}
 \rho_{DE}=\frac{1}{2k^2}J_{1},
 \end{equation}
 
\begin{equation}
 P=-\frac{1}{2k^2}\left(4J_{2}+J_{1}\right),
 \end{equation} where
 
\begin{equation}
 J_{1}=\left(-T-f+2TF\right),
 \end{equation}
 
\begin{equation}
 J_{2}=\left(1-F-2TF^{'}\right)\dot{H}.
 \end{equation} 
Here $F\equiv df/dT$, $F^{\prime}=dF/dT$. Transforming expressions (4), (5) and (7) we obtain
  
\begin{equation}
 -f+2TF=0
 \end{equation}   
hence

\begin{equation}
f=C\sqrt{T}.      
 \end{equation} 
 
Thus, we have obtained an exact solution for $f$, which satisfies as any Friedmann cosmology as for example a Little Rip model.
We take the Hubble parameter realizing Little Rip model as following [8]
  
\begin{equation}
 H=H_{LR}\exp(\xi{t}),
 \end{equation} 
where $H_{LR} (>0)$ and $\xi (>0)$ are positive constants. We see in (11) that in limit $t\rightarrow\infty$, $H$ diverges. In this case, the scale factor $a$ is expressed as
 
\begin{equation}
 a=a_{LR}\exp \left[ \frac{H_{LR}}{\xi}\exp\left(\xi{t}\right)\right],
 \end{equation} 
where $a_{LR}$ is a positive constant.

\section{Accelerating cosmology with dark matter} 
	Now, let us include the dark matter for this theory. In this case, the Friedman equation reads as
	
\begin{equation}
H^2=\frac{k^2}{3}\left(\rho_{DE}+\rho_{DM}\right),  
\end{equation}
where $\rho_{DM}$ is defined from continuity equation as
 
\begin{equation}
\rho_{DM}=a^{-3\left(1+\omega\right)}.
\end{equation}
Using expressions (4), (5) we get the equation for $f$ for accelerating cosmology with dark matter

\begin{equation}
-f+2TF=-2k^2\rho_{DM}.
\end{equation}
Here, choosing $f$ as in (10) gives us incoherent result. That's why as a form of $f(T)$, we take 

\begin{equation}
f=T^\alpha,
\end{equation}
where $\alpha(\neq0)$ is a non-zero constant. In this case, from Eq. (15) we have 

\begin{equation}
T^\alpha\left(1-2\alpha\right)=2k^2\rho_{DM}.
\end{equation}
Taking into account (14) and $T=-6H^2$, we obtain a scale factor $a$ depending from $t$ as

\begin{equation}
a=\left(\frac{2}{3}\frac{\alpha}{(1+\omega)\sqrt[2\alpha]{\frac{2k^2}{(-6)^\alpha(1-2\alpha)}}t}\right)^{-\frac{2\alpha}{3(1+\omega)}}. 
\end{equation}
This maybe considered as generalization of Little Rip cosmology in the presence of dark matter. It is seen that in the limit $t\rightarrow\infty$ when $\alpha>0$ and $\omega\geq0$ $a$ diverges, but in cases of $\alpha>0$, $\omega>0$ and $\alpha<0$, $\omega\geq0$ $a$ converges.     
 
\bigskip 

\section{Mimicking of $f(T)$ cosmology by inhomogeneous fluids}
	In this section we show that the same solutions obtained from $f(T)$ gravity can be taken in ordinary Einstein gravity with inhomogeneous dark fluid.  
	Let us rewrite a flat Friedmann-Lemaitre-Robertson-Walker space in equivalent form of the general relativity 
 
\begin{equation}
H^2=\frac{k^2}{3}\rho.
 \end{equation} 

 First, we consider Little Rip cosmology. For this model $\rho$ is defined from Eq. (19) as
\begin{equation}
 \rho=\frac{3}{k^2}H^{2}_{LR}\exp\left(2\xi{t}\right).
  \end{equation}

 We suppose that the universe is filled with an ideal fluid (dark energy) representing an inhomogeneous equation of state [9]
 
 \begin{equation}
 p=\omega(t)\rho+\Lambda(t),
 \end{equation}
where $\omega(t)$ and $\Lambda(t)$ are functions of time. 

The continuity equation is
 \begin{equation}
 \dot{\rho}+3H(\rho+p)=0.
 \end{equation}
Taking into account Eqs. (20)-(22) we obtain 

\begin{equation}
\rho\frac{2\xi}{3H_{LR}\exp(\xi{t})}+\rho\left(1+\omega(t)\right)+\Lambda(t)=0
\end{equation}

Assuming that effective cosmological constant and the density of dark energy connected in such a way $\Lambda(t)=\gamma\rho^2$, we find expression for $\omega$ from Eq. (23). Here $\gamma$ is a constant.

\begin{equation}
\omega(t)=-1-\frac{2\xi}{3H_{LR}\exp(\xi{t})}-\frac{3\gamma H^2_{LR}\exp(2\xi{t})}{k^2}.
\end{equation}
Substituting $p$ from Eq. (21), where $\omega$ and $\Lambda$ are defined as above, into Eq. (22) and using (19), (20) gives us the Hubble parameter of Little Rip model  

\begin{equation}
H=H_{LR}\exp(\xi{t}).
\end{equation}  
It is seen that Eqs. (19) and (22) are equivalent. Thus, we have shown that above inhomogeneous fluid realizes Little Rip, giving equivalence with $f(T)$ case [9].

Now, let us consider accelerating cosmology with dark matter. In this case, from expressions (18) and (19) the Hubble parameter takes a following form

\begin{equation}
H=\frac{2\alpha}{3(1+\omega)t}.
\end{equation}
Accordingly, the energy density for this cosmology is
\begin{equation}
\rho=\frac{4\alpha^2}{3k^2(1+\omega)^2t^2},
\end{equation}
Taking into accout (21), (22) and (27) we obtain

\begin{equation}
-\rho\frac{1+\omega}{\alpha}+\rho(1+\omega(t))+\Lambda(t)=0
\end{equation}
where 

\begin{equation}
\Lambda(t)=\gamma\rho^2=\frac{\gamma{16\alpha^4}}{9k^4(1+\omega)^4t^4}, 
\end{equation}
hence

\begin{equation}
\omega(t)=-1+\frac{1+\omega}{\alpha}-\frac{\gamma{4\alpha^2}}{3k^2(1+\omega)^2{t^2}}.
\end{equation}
By doing similar substitutions as in the Little Rip case, we get the same Hubble parameter for the accelerating cosmology with dark matter as given in Eq. (26). These two simple examples show that accelerating cosmology in $f(T)$ gravity maybe rewritten via equivalent description, in terms of general relativity with inhomogeneous fluid.  

\section{Examples of integrable $f(T)$ gravity models}
There are a very interesting class of $f(T)$ gravity models which is  integrable (see e.i. refs. \cite{14}-\cite{16}). Here we just present some of these models. To do that let us we assume that the function $f(T)$ is the solution of  Painlev$\acute{e}$ equations. 	These six Painlev$\acute{e}$ equations, traditionally called Painlev$\acute{e}$ I-VI equations or shortly, P-Z equations, where Z=I, II, III, IV, V, VI. They  are as follows:
\begin{equation}
 \begin{array}{|c|c|}
\hline \\
P_{I} - equation &   f^{\prime\prime}=6f^2+T  \\
\hline \\
P_{II} - equation &  f^{\prime\prime}=2f^3+T f+\alpha  \\
\hline \\
P_{III} - equation &   f^{\prime\prime}=\frac{1}{f} f^{\prime 2}-\frac{1}{T}(f^{\prime}-\alpha f^2-\beta)+\gamma f^3+\frac{\delta}{f} \\
\hline \\
P_{IV} - equation &   f^{\prime\prime}=\frac{1}{2f}f^{\prime 2}+1.5f^3+4T f^2+2(T^2-\alpha)f+
 \frac{\delta}{f} \\
\hline \\
P_{V} - equation &   f^{\prime\prime}=(\frac{1}{2f}+\frac{1}{f-1})f^{\prime 2}
 -\frac{1}{T}(f^{\prime}-\gamma  f)+\frac{(f-1)^2}{T^{2}}(\alpha f+\frac{\beta}{f})
 +\frac{\delta f(f+1)}{f-1} \\
\hline \\
P_{VI} - equation &  f^{\prime\prime}=\varphi(T)f^{\prime 2}
 -\xi(T)f^{\prime}
 +\frac{f(f-1)(f-T)}{T^{2}(T-1)^{2}}[\alpha+ \frac{\beta T}{ f^{2}}+\zeta(T)]\\
\hline
\end{array} \end{equation}
Here $\varphi(z)=0.5[f^{-1}+(f-1)^{-1}+(f-z)^{-1}], \quad \xi(z)= [z^{-1}+(z-1)^{-1}+(f-z)^{-1}],\quad \zeta(z)=[\gamma(z-1)(f-1)^{-2}+\delta z(z-1)(f-z)^{-2}]$. All of these $P_{I}$-$P_{VI}$ models (equations) admit  infinite number integrals of motions, $n$-soliton solutions, Lax representations and so on. These integrable $f(T)$ gravity models we will study in detail separately.
Finally we note that also it is interesting to extend the above obtained results for the $F(R,T)$ gravity, where $R$ is the Ricci scalar and $T$ is the torsion scalar \cite{17}-\cite{23}. 

\bigskip

\section{Conclusion}
	In summary, using the cosmological solutions obtained in $f(T)$ gravity, we have shown that the same solutions can be obtained through effective inhomogeneous dark fluid. Here, the inhomogeneous dark fluid leads to effects equivalent to teleparallelism. The obvious question is how to allocate the physical scenario of accelerated expansion of the Universe among several equivalent theoretical scenarios? In this case, it is $f(T)$ cosmology and cosmology with inhomogeneous dark fluid. It is clear that future observational tests (for example, cosmography [10]) should take into account this problem and propose its solutions.


\end{document}